# Experimental Study of Decentralized Robot Network Coordination

Martyn Lemon and Yongqiang Wang

Department of Electrical and Computer Engineering, Clemson University

Clemson, SC 29634, USA

**Synchronization and desynchronization in networks is a highly studied topic in many electrical systems, but there is a distinct lack of research on this topic with respect to robotics. Creating an effective decentralized synchronization algorithm for a robotic network would allow multiple robots to work together to achieve a task and would be able to adapt to the addition or loss of robots in real-time. The purpose of this study is to improve algorithms implemented developed by the authors for this purpose and experimentally evaluate these methods. The most effective algorithm for synchronization and desynchronization found in a former study were modified to improve testing and vary its methods of calculation. A multi-robot platform composed of multiple Roomba robots was used in the experimental study. Observation of data showed how adjusting parameters of the algorithms affected both the time to reach a desired state of synchronization or desynchronization and how the network maintained this state. Testing three different methods on each algorithm showed differing results. Future work in cooperative robotics will likely see success using these algorithms to accomplish a variety of tasks.**

## I. INTRODUCTION

The purpose of this paper is to study navigation and synchronization techniques for robot networks using microcomputers as controllers. It is also to study the use of IMU integrated circuits to gather readings on acceleration, gyration and magnetic field and use these to accomplish various tasks. To study synchronization and communication, radio frequency transmitters are used. Studying how to use these radio frequency devices to communicate in an effective way and how to write programs that allow multiple Roomba robots to run code simultaneously was a central focus of this paper. The primary goal was to improve and test different methods of the synchronization and desynchronization algorithms developed previously by the authors [1-6]. These methods, which will be discussed in further detail, altered the way the robots sought to achieve the desired state of communication. The ultimate goal of this research would be to extend synchronous robotic networks to useful industries. Search and rescue, agriculture, the military and logistics are all industries that could benefit from multiple robots being able to work together to accomplish a task. By making the network decentralized, the hope is that one robot malfunctioning would not cause the network to fail, rather the remaining robots could readjust to attain the same state as before.

## II. METHODOLOGY

As described before, the platform used to test the synchronization algorithms was on the Roomba$^{TM}$, specifically a modified version with the vacuum brushes removed. To interface with the I/O port that was connected to the wheels and sensors, a Raspberry Pi 3 Model B microcomputer was used. This controller is WiFi enabled, so programs could be transferred to it wirelessly and the Raspbian operating system could be accessed using SSH from a laptop. The IMU modules were mounted directly on the Roomba using a mounting bracket developed and 3D printed by other engineers in the lab, shown in Figure 1 below. Figure 2 shows the Xbee module, a radio frequency transmitter connected via USB to the Raspberry Pi. This is what we used to make the synchronization happen.

To create these algorithms, the Python language was used. The two algorithms tested this semester were the PRC method of synchronization and the PRC method of desynchronization.. The first addition to these programs was functionality that allowed the user to specify an initial angle for the Roomba, rather than calibrating it using the magnetometer. Since the tests were conducted indoors in an area with high magnetic interference, the readings given by these sensors were often very inaccurate. Since the starting angle of each Roomba was so inconsistent across tests, this skewed some of the previous data. By setting each Roomba to be equally spaced apart for a synchronization test or to the same angle for desynchronization, we ensure a maximum workload for the algorithm.

Additional changes included the new methods by which the algorithms operated. The optimized spin method was made so that while moving, the Roomba would calculate the magnitude of its desired turn first, and then the direction, rather than doing this simultaneously. The goal here was to observe how this affected the speed to achieve steady state and how it affected that steady state. The other two methods created were the constant time method and the constant frequency method [1-7]. The constant time method works by ensuring all Roombas will be turning at the same and for the same amount of time, though they can have different speeds. The constant frequency method works conversely—each Roomba can spin for a different length of time, but they will always turn at the same rate.

One of the new implementations for the project was using Github for file storage, transfer and editing. Creating a Github repository and accessing it from the Raspberry Pi computers allowed a centralized place for everyone in the lab to add, share and edit files in one location accessible on any computer. Since each test for the algorithms for this part of the project created six large text files, one on each Roomba, being able to transfer these directly from the Raspberry Pi to the central database made work much more efficient. Learning and understanding the UNIX commands to interact with Github was challenging at first, but after a few sessions of testing they became much more fluid. Additionally, the code to analyze the data and create graphs in Matlab was reworked to be able to handle six datasets at once—prior to this semester the most tested at one time was four.

### III. Data

The graphs shown below are the test results of each different test conducted this semester. The so-called "containing arc" is a measurement that is the total difference from the desired state of either synchronization or desynchronization. The Matlab code prompts the user if the test was either using the sync or desync code and will calculate this metric accordingly. The closer the arc is to zero, the better the results of the test are.

Figures 3, 4 and 5 are the synchronization tests. The first test was using the optimized spin method, the second was the constant time method and the third the constant frequency method. It is easily apparent that while the frequency method is effective at getting to the desired state, it has much more oscillatory behavior than the other two. For the desynchronization tests, a similar pattern emerges. Figure 7 and 8 are both using the optimized spin method—7 has higher coefficients applied to the turn values than 8. Figure 9 shows the constant time method, and 10 shows the frequency method.

### IV. Limitations

There were some constraints that prevented ideal data from being collected. The Roombas were sometimes inconsistent and would fail unexpectedly during a test, so this proved frustrating since the Matlab code would have to be readjusted to work for five Roombas instead of six. However, it was interesting to note that if one did fail, the others could still find and achieve the desired state—the Roombas would readjust accordingly. Another limitation was having to specify the angle of the Roomba before rather than using the magnetometer. Even though specifying the angle made the ideal condition for the algorithm to work. It would have been interesting to observe if using the magnetometer for angle measurement rather than a pre-defined value and the wheel encoders would make a substantial difference on the efficacy of these algorithms.

Another limitation was the time required to set up and conduct each test. Ideally, more data could have been collected varying parameters within the algorithm, and the new methods developed this semester could have been tested on older algorithms to see if they would make them work any better. Future research in this topic might see this happen.

### V. Results and Conclusions

An analysis of this data reveals that the optimized spin method [1] was the most effective for desynchronization. It reached the desired behavior in the least amount of cycles, and its steady-state oscillations were minimal compared to other methods. For the synchronization, it appears that the optimized spin method and the constant time method were similarly effective, while the constant frequency method was significantly less effective, taking longer and creating more oscillations. Future work should see more tests conducted using differing parameters for these methods. Additionally, future research will ideally involve making code that will allow the Roombas to perform more complex tasks using this state of synchronization. As of now, all they do is turn to either the same angle or a perfectly desynchronized combination of angles. Perhaps an easy thing to implement would be for the keyboard control to move the Roombas around. Having this work synchronously would allow a user to control a group of Roombas and ensure they would all move the same way.

In the future, this work could be extended to other robotic platforms and hopefully onto more complex tasks. Implementing this work with other research in the lab would be ideal as well. The principal work of other researchers in the lab was on navigation for the Roombas and mapping out a two-dimensional floor space to make navigation more efficient. Being able to synchronize multiple Roombas and have them collectively map out an area would make the process much more efficient.

### VI. Figures

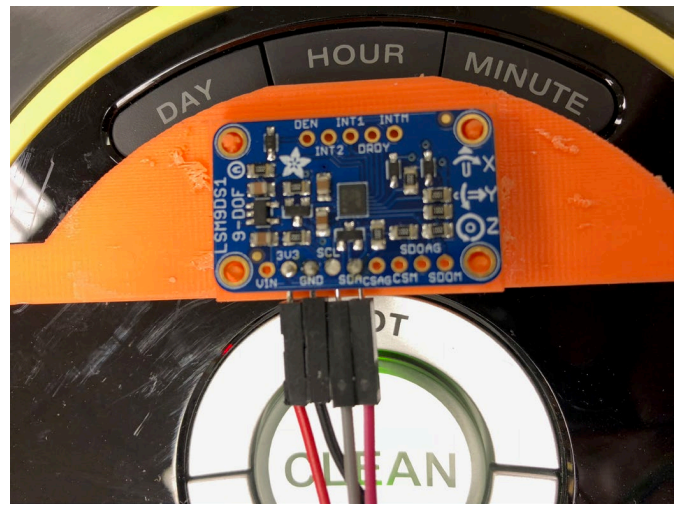

**Figure 1**

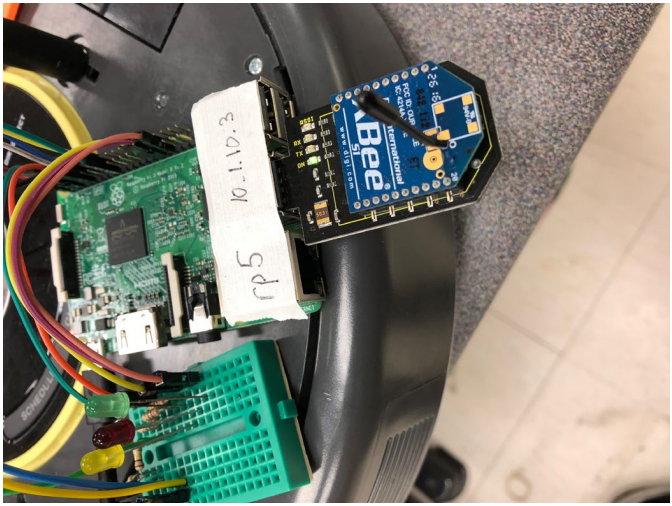

Figure 2

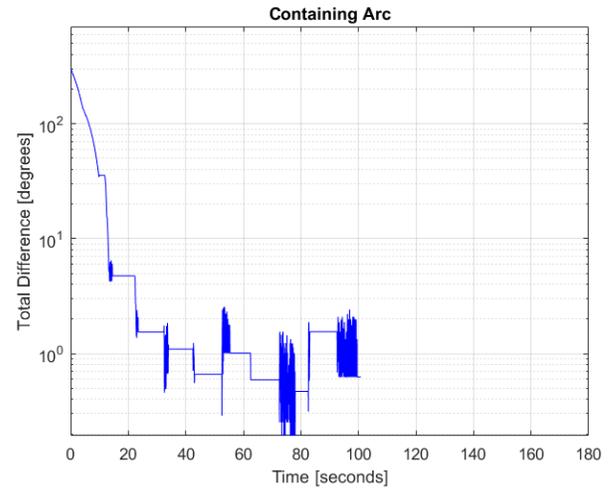

Figure 5

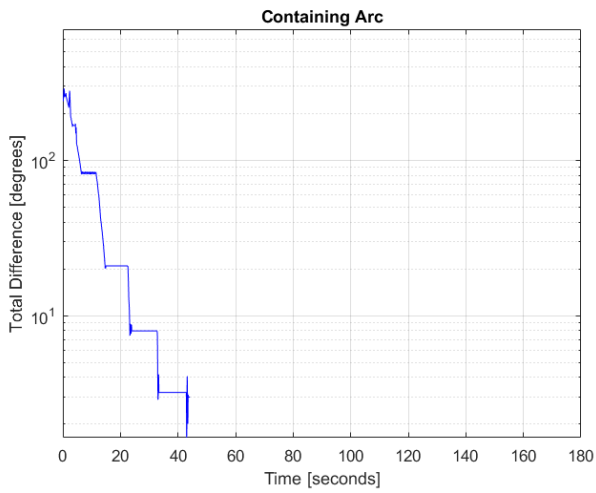

Figure 3

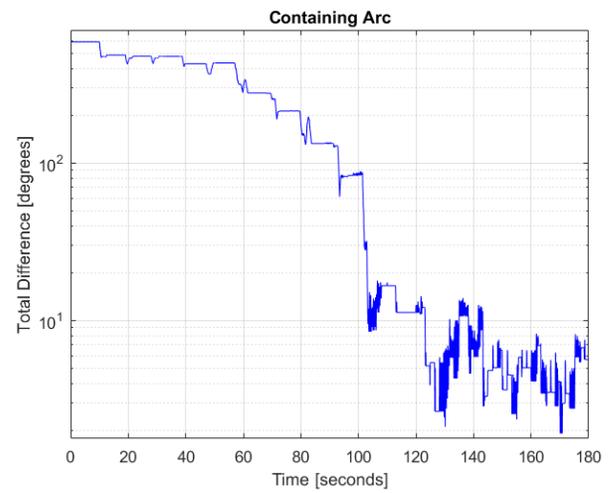

Figure 6

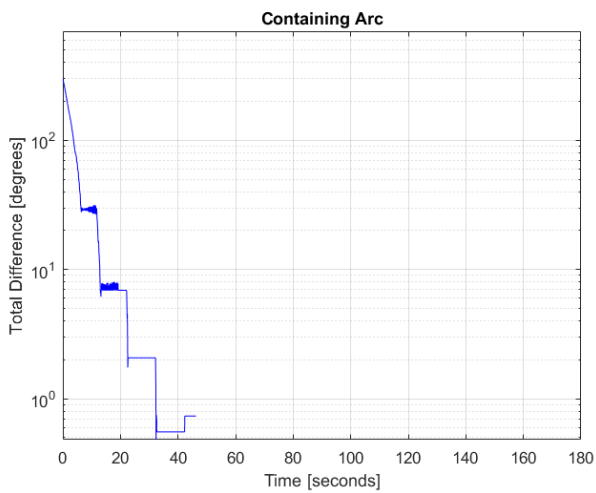

FIGURE 4

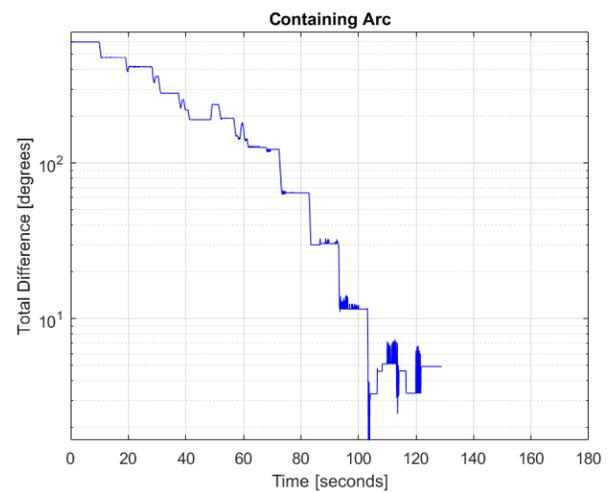

FIGURE 7

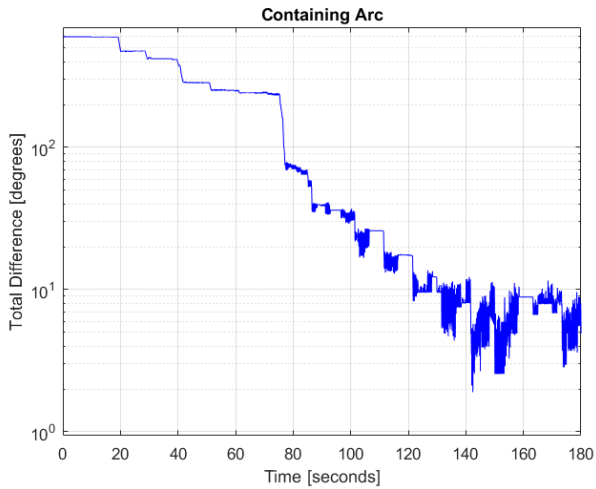

**Figure 8**

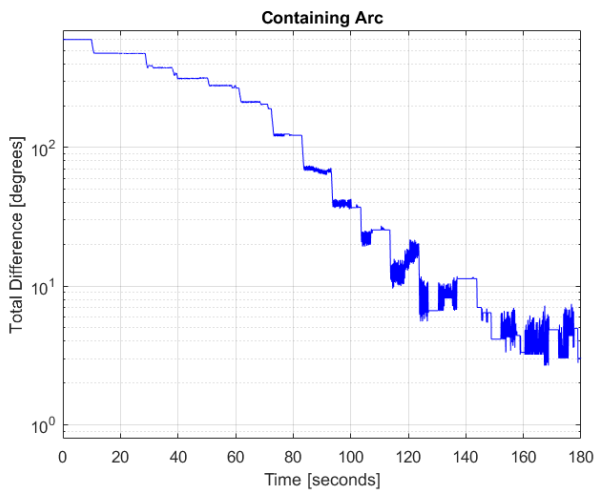

**Figure 9**

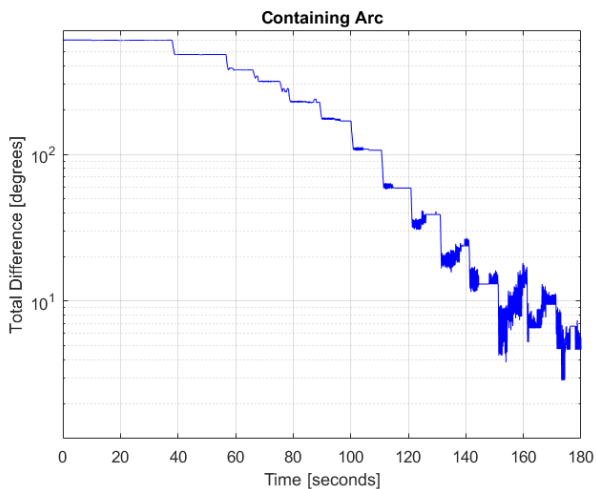

**Figure 10**

ACKNOWLEDGMENT

Clemson University, the Department of Electrical and Computer Engineering, Dr. Yongqiang Wang and Mr. Timothy Anglea are all thanked for their invaluable support of this project.

REFERENCES

[1] T. Anglea and Y. Q. Wang, "Synchronization with guaranteed clock continuity using pulse-coupled oscillators." [Online]. Available: https://arxiv.org/pdf/1704.07201.pdf

[2] Gao H, Wang Y Q. A Pulse Based Integrated Communication and Control Design for Decentralized Collective Motion Coordination. IEEE Transactions on Automatic Control. 2018, 63(6): 1858-1864

[3] Ferrante F, Wang Y Q. Robust almost global splay state stabilization of pulse coupled oscillators. IEEE Transactions on Automatic Control, 2017, 62(6): 3083 – 3090.

[4] Gao H, Wang Y Q. On Phase Response Function based Decentralized Phase Desynchronization. IEEE Transactions on Signal Processing, 2017, 65(21): 5564 – 5577.

[5] Anglea T, Wang Y Q. Phase desynchronization: a new approach and theory using pulse-based interaction. IEEE Transactions on Signal Processing, 2017, 65(5):1160-1171.

[6] Wang Y Q, Nunez F, Doyle F. Energy-efficient pulse-coupled synchronization strategy design for wireless sensor networks through reduced idle listening. IEEE Transactions on Signal Processing, 2012, 60(10):5293-5306.

[7] Wang Y Q, Ye H, Ding S X, Wang G Z. Fault detection of networked control systems based on optimal robust fault detection filter. Acta Automatica Sinica, 2008, 34 (12), 1534-1539

[8] Wang Y Q, Núñez F, and Doyle F. Statistical analysis of the pulse-coupled synchronization strategy for wireless sensor networks. IEEE Transactions on Signal Processing, 2013, 61 (21): 5193-5204.

[9] Wang Y Q, Doyle F. J. On influences of global and local cues on the rate of synchronization of oscillator networks. Automatica, 2013, 47(6): 1236-1242.

[10] Zhang, Kaixiang, et al. Privacy-preserving dynamic average consensus via state decomposition: Case study on multi-robot formation control. Automatica, 139 (2022): 110182.